\documentclass[showpacs,pra,twocolumn]{revtex4}
\usepackage{amsmath}
\usepackage{amsfonts}
\usepackage{graphicx}
\usepackage{dcolumn}
\usepackage{stackrel,amssymb}
\usepackage{color}
\usepackage{ulem}
\usepackage{float}

\usepackage{bm}

\begin{document}

\title{Probing the spectral density of a dissipative qubit via quantum synchronization}

\author{Gian Luca Giorgi, Fernando Galve, and Roberta Zambrini}
\affiliation{ Instituto de
F\'{\i}sica Interdisciplinar y Sistemas Complejos IFISC (UIB-CSIC), UIB Campus,
E-07122 Palma de Mallorca, Spain}

\begin{abstract}
The interaction of a quantum system, which is not accessible by direct measurement, with an external  probe can be exploited to infer specific features of the system itself. 
We introduce a probing scheme based on the emergence of spontaneous quantum synchronization
between an out-of-equilibrium qubit, in contact with an external environment, and a probe qubit. 
Tuning the  frequency of the probe leads to a transition between synchronization
in phase and  antiphase. The sharp transition between these two regimes is locally accessible by monitoring the probe dynamics alone and allows one to reconstruct the shape of the  spectral density of the environment. 
\end{abstract}


\maketitle

\section{Introduction}

Probing individual quantum systems interacting with their environments is
instrumental for their exploitation and control in many applications.
Starting from the 1990s, different strategies to probe the dynamical features of
microscopic systems have been implemented that infer oscillation frequency and dissipation at the atomic level and in nanodevices either optically or electrically \cite{Itano,Jelezko,Loth}.

A probe is a physical object interacting with a hardly accessible or measurable complex system 
in  such a way that, monitoring the probe itself, it is possible to extract information about the system.
Proposed schemes include, but  are not limited to, probing of
temperature \cite{temp}, work distribution \cite{work}, non-Markovianity
\cite{nm}, Hamiltonian tomography \cite{ht}, phase transitions \cite{pt}, 
excitation spectra \cite{es}, and spectral densities \cite{eisert}. 

Most techniques are hybrid, with  the probe and the system being different in
their physical natures. Furthermore, the probes are usually classical objects  (for instance,  laser beams). 
An alternative approach consists of coupling two similar units,
 one of them monitoring the other one. This happens, for
  example, in the case of an ion (the system to be probed)
interacting with an auxiliary one confined in a trap, which serves as a probe because it is more favorable for precision spectroscopy \cite{Schmidt}.
As a  second example,  a boson (whose dynamics is recorded) is used to probe the structure of a complex bosonic  network   \cite{nokkala}.  
Indeed, the key question is to establish which features of the system can be extracted accessing only the probe
and how favorable a scheme is.

In this paper, we propose to probe the evolution of a single, dissipative, out-of-equilibrium qubit through a second
detuned qubit, exploiting the phenomenon of spontaneous synchronization (SS).   
This is a paradigmatic and widely explored phenomenon in physical, biological,
chemical and 
social contexts \cite{Strogatz,Pikovsky,Arenas}, whose extension to the quantum
world has become the
object of extensive studies during the last few years.
The initial approaches dealt with synchronization as a response to an external driving field, a phenomenon known as entrainment \cite{entrain}. These studies were followed by the investigation of quantum SS in mechanical resonators \cite{mec}, 
harmonic networks \cite{PRAsync,syncNET}, coupled ultracold atomic gases
\cite{julia},   van der Pol  oscillators \cite{vdp}, cold ions in microtraps
\cite{hush}, ensembles of quantum dipoles \cite{zhu}, and uncoupled spin systems
dissipative through a common environment \cite{Giorgi}. 

The phenomenon of quantum SS is of twofold interest: 
on the one hand, it
naturally arises in extended systems due to the interaction between their components
that leads to collective coherence, and on the other hand, it can serve as a
tool in different applications, an avenue open to exploration with important antecedents
for classical SS \cite{Chaos}. Here, we demonstrate quantum
SS in a system-probe setup (widening what was found in 
previous works \cite{syncNET,Giorgi,leHur}) with an interesting abrupt transition
between synchronization in phase and  antiphase.  This  is exploited in 
a novel quantum probing scheme where this sharp instability enables
the  reconstruction of the dissipative qubit spectral density.

\section{Model}

\begin{figure}[t]
\includegraphics[width=6cm]{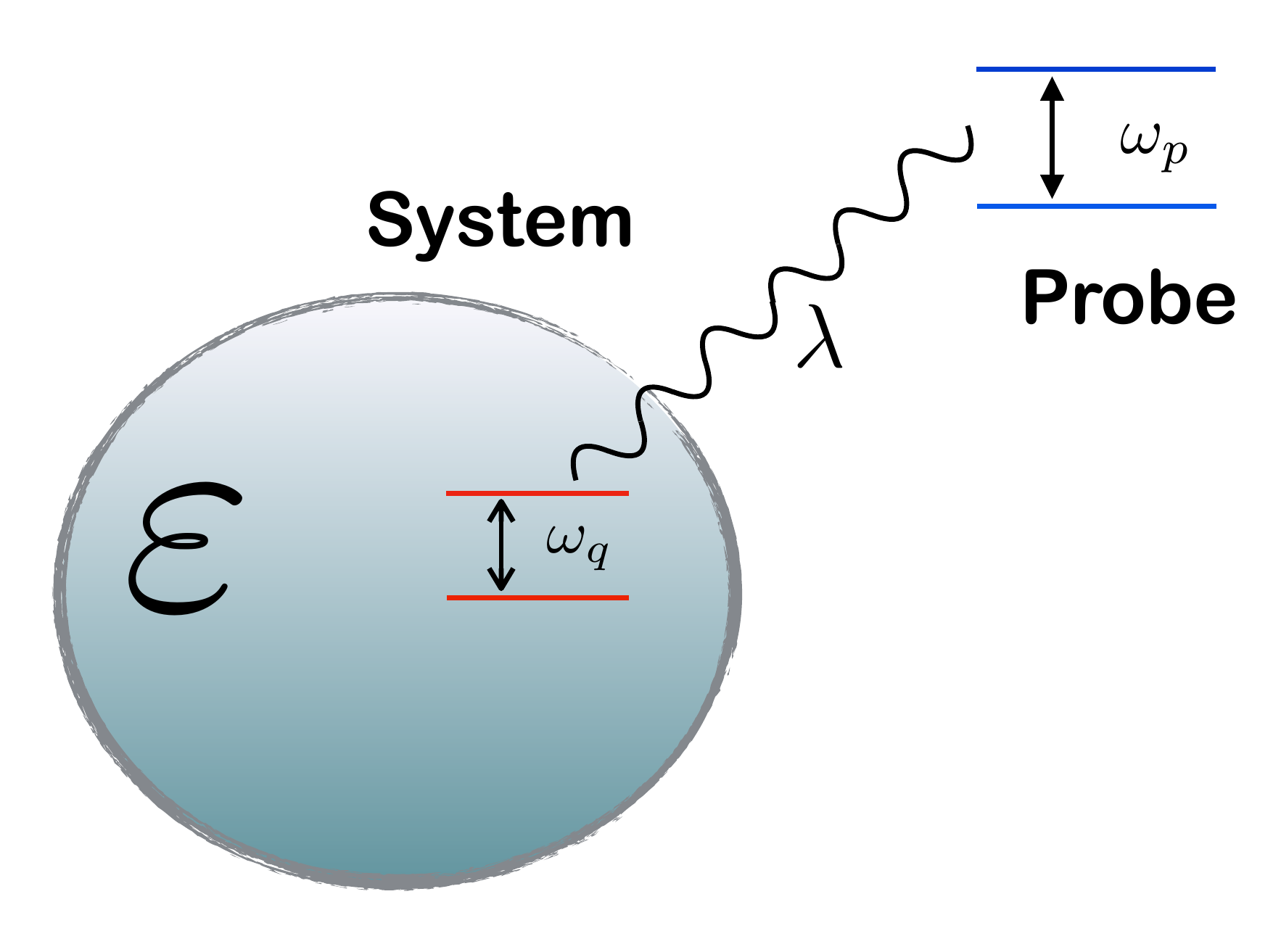}
\caption{Schematics of the model: a system $q$, interacting with its own thermal environment, is connected to an external probe $p$.}\label{figschematics}
\end{figure}

Let us consider a qubit $q$ immersed in an external dissipative environment  
with the microscopic Hamiltonian  
\begin{equation}
H_0=\frac{\omega_q}{2}\sigma_q^z+\sum_k \Omega_k a_k^\dag a_k +\sum_k g_k (a_k^\dag+ a_k )\sigma_q^x,
\end{equation}
with $\sigma_q^i$ ($i=x,y,z$)  being Pauli matrices and  bosonic  bath eigenmodes  $a_k$  with energies $\Omega_k$
($\hbar$ is set to $1$ throughout the paper).
The dissipative process is fully determined by the spectral density $J(\omega)= \sum_k g^2_k\delta(\omega-\Omega_k)$.
Is it possible to reconstruct the features of this dissipative qubit by coupling it with another qubit (a quantum probe $p$)
instead of directly measuring it?  We assume that the system and the probe interact through an Ising-like coupling. Then, the total Hamiltonian is
\begin{equation}
H=H_0+\frac{\omega_p}{2}\sigma_p^z+\lambda \sigma_q^x \sigma_p^x. \label{htot}
\end{equation}
This kind of interaction has been  experimentally reported in different physical contexts \cite{friedenauer,simon,viehmann,yan}.
The extension to the more general class of anisotropic  $XY$ coupling would give similar results.
A schematic representation of the model is presented in Fig. \ref{figschematics}.

We point out that even though the environment directly interacts only   with the system qubit $q$,   the effects over the full qubit-probe system $H_S=\frac{\omega_q}{2}\sigma_q^z+\frac{\omega_p}{2}\sigma_p^z+\lambda\sigma_q^x \sigma_p^x$ need to be accounted for in order to work out a master equation valid for any system-probe coupling strength $\lambda$.

Assuming weak dissipation, the   dynamics of the pair of qubits can be studied in the Born-Markov 
and secular approximations 
with the Lindblad master equation $\dot\rho(t)=-i [H_{S}+H_{LS},\rho(t) ]+{\cal D}[\rho(t)]$, where the
Lamb shift  $H_{LS}$ commutes with $H_{S}$ 
and  ${\cal D}[\rho(t)]$ is the standard dissipator   \cite{bp}.

In order to calculate $\rho(t)$, we diagonalize  $H_S$ through the standard Jordan-Wigner technique, 
which maps spins into spinless fermions \cite{lsm}. 
This leads to (see Appendix \ref{appA}) $H_S=E_1 (\eta_1^\dag \eta_1-1/2)+E_2 (\eta_2^\dag \eta_2-1/2)$,
with  
\begin{eqnarray}
 2E_1&=&\sqrt{4\lambda^2+\omega_+^2}+\sqrt{4\lambda^2+\omega_-^2},\\
 2E_2&=&\sqrt{4\lambda^2+\omega_+^2}-\sqrt{4\lambda^2+\omega_-^2},
 \end{eqnarray}
where $\omega_\pm=\omega_q\pm \omega_p$ 
and  the quasiparticle vacuum $|0,0\rangle$ is the ground state.
 In terms of fermions, the  operator $\sigma_q^x$ 
 appearing in the system-bath Hamiltonian becomes 
 \begin{equation}
 \sigma_q^x =\cos (\theta_++\theta_- )(\eta_1^\dag+ \eta_1)+\sin(\theta_++\theta_-)(\eta_2^\dag+ \eta_2),\label{sq}
 \end{equation}
 where $\theta_\pm=\arcsin(2 \lambda/\sqrt{4\lambda^2+\omega_\pm^2})/2$.
Under the secular approximation, the dissipative part of the master equation is
\begin{equation}
{\cal D}(\rho)=\sum_{i=1}^2\tilde\gamma_i^+ {\cal L}[\eta_i](\rho)+\sum_{i=1}^2\tilde\gamma_i^- {\cal L}[\eta_i^\dag](\rho),\label{dissip}
\end{equation}
 with $\tilde\gamma_1^+=\cos^2 (\theta_++\theta_-)J( E_1)[1+n(E_1)]$, $\tilde\gamma_1^-=\cos^2 (\theta_++\theta_-)J( E_1)$, $\tilde\gamma_2^+=\sin^2 (\theta_++\theta_-)J( E_2)[1+n(E_2)]$, and $\tilde\gamma_2^-=\sin^2 (\theta_++\theta_-)J( E_2)$, where $n(.)$ stands for the Bose factor and  the Lindblad superoperator is, as usual, 
$ {\cal L}[\hat X](\rho)=\hat X \rho \hat X ^\dagger -\{\hat \rho ,\hat X^\dagger \hat X \}/2$. We stress that the present master description is also valid for strong 
coupling $\lambda$ between probe and system qubits (as long as the system-bath coupling is kept small) and that we consider generic  initial conditions (in contrast to Refs. \cite{Campbell} and \cite{Jeske}). In this broader scenario, dynamical SS emerges between detuned qubits \cite{footnote} as described qualitatively and analytically in the following.

\section{Quantum synchronization}

\begin{figure}[t]
\includegraphics[width=8cm]{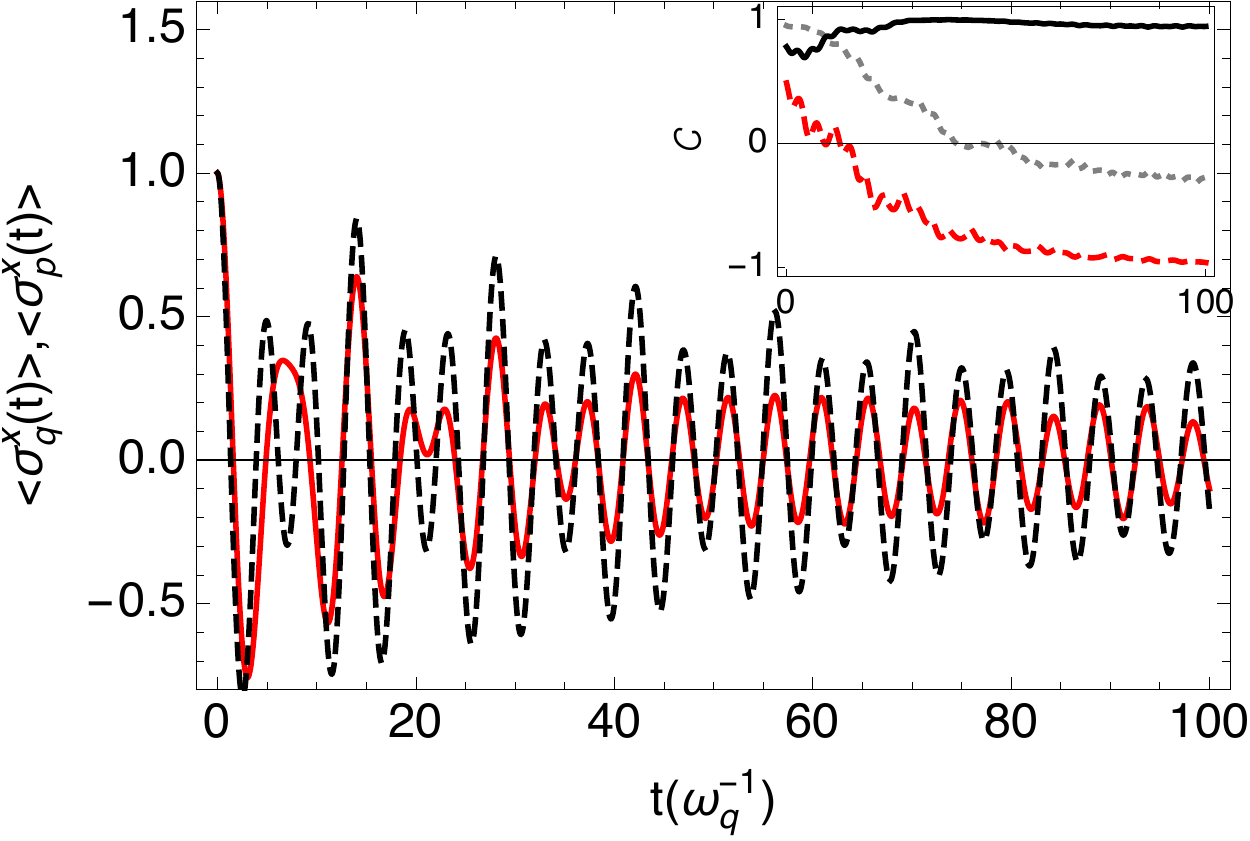}
\caption{Dynamical synchronization of $\langle\sigma_q^x\rangle$ (red  solid line)
and $\langle\sigma_p^x\rangle$ (black  dashed line) for $\omega_p=1.2\,\omega_q$ and 
$\lambda=0.2\,\omega_q$. The bath spectral
density is Ohmic:  $J(\omega)=2 \,\gamma_0\, \omega
\,\omega_c^2/(\omega_c^2+\omega^2)$ 
with  cut-off frequency  
$\omega_c=20\, \omega_q$ and $\gamma_0=10^{-2}\omega_q$ and $T=0$.  The initial state is
$|\psi(0)\rangle=(|0\rangle+|1\rangle)(|0\rangle+|1\rangle)/2$. 
The inset shows the synchronization measure ${\cal C}_{\sigma_p^x,\sigma_q^x}(t)$  for 
 $\omega_p/\omega_q=1.2$ (black solid line), $\omega_p/\omega_q=0.8$ (red dashed line), and $\omega_p/\omega_q=1$ (gray dotted line),
 as examples of SS in phase and antiphase and the absence of SS, respectively.  }
\label{fig1}
\end{figure}

The emergence of dissipation-induced quantum SS has been discussed
in the cases of bosons and noninteracting spins in the presence of a common environment \cite{PRAsync,syncNET,Giorgi}. The model we are discussing
here shows a different scenario  as SS between the system qubit
and the probe is predicted in spite of the absence of a common bath (see Fig.
\ref{fig1}).

In the presence of competition between different oscillating modes, synchronization is achieved whenever there is appreciable
separation between the two largest 
decay times  characterizing the dynamics. Then,
slowly decaying local degrees of freedom   experience monochromatic oscillations
at the same surviving frequency, 
while the relative phases among them are locked (see Appendix \ref{appC}).
The argument can be made quantitative by noticing that any observable $O$ 
can be decomposed  in terms of its frequency jumps:  $O=\sum_\omega O(\omega)$ \cite{bp}, where $
O(\omega)=\sum_{\varepsilon}\Pi(\varepsilon) O\; \Pi(\varepsilon+\omega),
$
and  $\varepsilon$ is the set of eigenvalues of $H_S$, while
$\Pi(\varepsilon)$ is the projection onto the corresponding eigenspace. In the
presence of degeneracy (which is the case 
under study, as there are pairs of transitions at energies $E_1$ and $E_2$) the time evolution in the Heisenberg picture takes the general form
$ \langle O(t)\rangle=\sum_{k,\omega} \langle O_k(\omega)\rangle e^{-i \omega t}e^{-\Gamma_k(\omega)t}$,
where the index $k$ spans the degeneracy subset. The normal operators
$  O_k(\omega)$ are obtained by diagonalizing the dynamical  equations of motion
within the degeneracy subspace.
Taking any couple of local observables $O^{(1)}$ and $O^{(2)}$, they will
experience synchronization if the smallest nonzero decaying rate $\Gamma_{\bar
k}$ is ``seen" by both of them and, 
at the same time, is much smaller than any other $\Gamma_k$. While the presence
of two  separate, identical environments may hinder
synchronization 
\cite{PRAsync}, here,
the interaction with one local environment favors SS for almost any choice
of the system's parameters; therefore, this phenomenon is a robust feature of this model.  
Indeed, in the proposed setup, $strongly$ detuning the two spins favors the emergence of SS,
as it causes an imbalance among the couplings of the eigenmodes of $H_S$ to the bath and then a marked 
separation of the  $\Gamma_k$.
This peculiar effect is desirable as, otherwise, ascertaining the proper tuning of the probe  would  be 
problematic for unknown system frequency.

The local variables we monitor are $\langle\sigma_q^x(t)\rangle$  and  $\langle\sigma_p^x(t)\rangle$,
whose decomposition in terms of fermionic quasiparticles is given by Eq. (\ref{sq}) and by
\begin{equation}
\sigma_p^x =\sin (\theta_+-\theta_- )(\tilde\eta_1^\dag+ \tilde\eta_1)+\cos(\theta_+-\theta_-)(\tilde\eta_2^\dag+\tilde\eta_2),
\end{equation}
with $\tilde\eta_i={\cal P}\eta_i$, where ${\cal P}=(1-2 \eta_1^\dag \eta_1)(1-2 \eta_2^\dag \eta_2)$ is the parity operator.

Due to the form of the dissipator, the quasiparticle operators in the interaction picture obey 
\begin{eqnarray}
\langle\dot{\eta_i}\rangle&=&-\frac{1}{2}(\tilde\gamma_i^+ +\tilde\gamma_i^-+2\tilde\gamma_j^+ +2\tilde\gamma_j^-)\langle\eta_i\rangle-(\tilde\gamma_j^- -\tilde\gamma_j^+)\langle\tilde{\eta}_i\rangle,\nonumber\\
\langle\dot{\tilde{\eta}}_i\rangle&=&-\frac{1}{2}(\tilde\gamma_i^+ +\tilde\gamma_i^-)\langle\tilde{\eta}_i\rangle,\label{eqm}
\end{eqnarray}
with $i,j=1,2$ and $i\neq j$. Then, after a transient, we have 
$\langle\tilde{\eta}_i(t)\rangle=\langle\tilde{\eta}_i(0)\rangle\exp{[-(\tilde\gamma_i^+ +\tilde\gamma_i^-) t/2]}$ 
together with $\langle\eta_i(t) \rangle\sim\langle\tilde{\eta}_i(t)\rangle\times (\tilde\gamma_j^- -\tilde\gamma_j^+)/(\tilde\gamma_j^- +\tilde\gamma_j^+)$. 
The evolutions simplify in the zero-temperature limit, 
when $\tilde \gamma_i^-=0$, as in this case 
\begin{eqnarray}
\langle\sigma_q^x(t)\rangle &\sim & 2 \cos( \theta_++ \theta_-)e^{-\tilde\gamma_1^+t/2}
  {\rm Re}[e^{i E_1 t }\langle\tilde{\eta}_1(0)\rangle ]\nonumber\\
&+&2\sin(\theta_++\theta_-)e^{-\tilde\gamma_2^+t/2} {\rm Re}[e^{i E_2 t }\langle\tilde{\eta}_2(0)\rangle ] \label{sq1},\\
\langle\sigma_p^x(t)\rangle &\sim & 2 \sin( \theta_+ - \theta_-)e^{-\tilde\gamma_1^+t/2}
  {\rm Re}[e^{i E_1 t }\langle\tilde{\eta}_1(0)\rangle ]\nonumber\\
&+&2\cos(\theta_+-\theta_-)e^{-\tilde\gamma_2^+t/2} {\rm Re}[e^{i E_2 t }\langle\tilde{\eta}_2(0) \rangle]. \label{sp1}
\end{eqnarray}
Then,  synchronization takes place whenever $\tilde\gamma_1 \ll \tilde\gamma_2$
or  $\tilde\gamma_1 \gg \tilde\gamma_2$. Apart from the special cases
where the two decaying rates are of the same order of magnitude, 
after the faster mode has decayed, $\langle\sigma_q^x\rangle$ and
$\langle\sigma_p^x\rangle$ oscillate at the same frequency $\omega_{{\rm
sync}}$, with the synchronization frequency being close to either $E_1$ or $E_2$ 
(up to the negligible Lamb shift). It turns out that when  $\omega_{{\rm sync}}\simeq
E_1$, the two spins are anti-synchronized, while they are 
synchronized when $\omega_{{\rm sync}}  \simeq  E_2$. This is due to the sign of
the ratio between  the trigonometric prefactors  entering Eqs. 
(\ref{sq1}) and (\ref{sp1}).  Indeed,   up to unobservable rotations, 
$0\le \theta_+\le \theta_-\le \pi/4$. Then, we always have $\cos( \theta_++
\theta_-)/\sin( \theta_+ - \theta_-)\le 0$ and $\sin( \theta_+ + \theta_-)/\cos(
\theta_+ - \theta_-)\ge 0$. The analysis is similar for temperature $T\neq 0$, as detailed in   Appendix \ref{appC}. 
 All the qualitative description given so far still holds. However, we must take into account that, as all the decay rates become faster, for very high temperature the dissipation can become so fast with respect to the  frequency that it may become impossible to observe any oscillation. Then. a natural limit to observe synchronization is obtained by comparing the smallest rate to the number of cycles needed to actually observe SS. An example is given in Appendix \ref{appC}.

Synchronization between  the two local observables  $\langle\sigma_{p,q}^x(t)\rangle$ can be quantified by looking at their
normalized time correlation   ${\cal C}$ \cite{PRAsync,Giorgi} (alternative approaches \cite{ameri,zhu} will be discussed later).  Given two time-dependent functions $f$ and $g$, it is defined as
\begin{equation}
{\cal C}_{f,g}(t,\Delta t)=\frac{\overline{\delta f\delta g}}{\sqrt{\overline{\delta f^2}\; \overline{\delta g^2}}},
\end{equation}
where the bar stands for the time average 
\begin{equation}
\overline{f}=\frac{1}{\Delta t}\int_t^{t+\Delta t} f(t^\prime)d t^\prime
\end{equation}
performed over the time window $\Delta t$ and $\delta f=f-\overline{f}$.
 An absolute value of ${\cal C}$ close to $1$ would indicate a high degree of
synchronization, 
while indicates ${\cal C}\sim 0$ the absence of synchronization. In the  inset of Fig.
\ref{fig1}, we show both synchronized and antisynchronized
regimes as well as the absence of SS by varying $\omega_p$.

\section{Synchronization as a probing tool}

\begin{figure}[t]
\includegraphics[width=8cm]{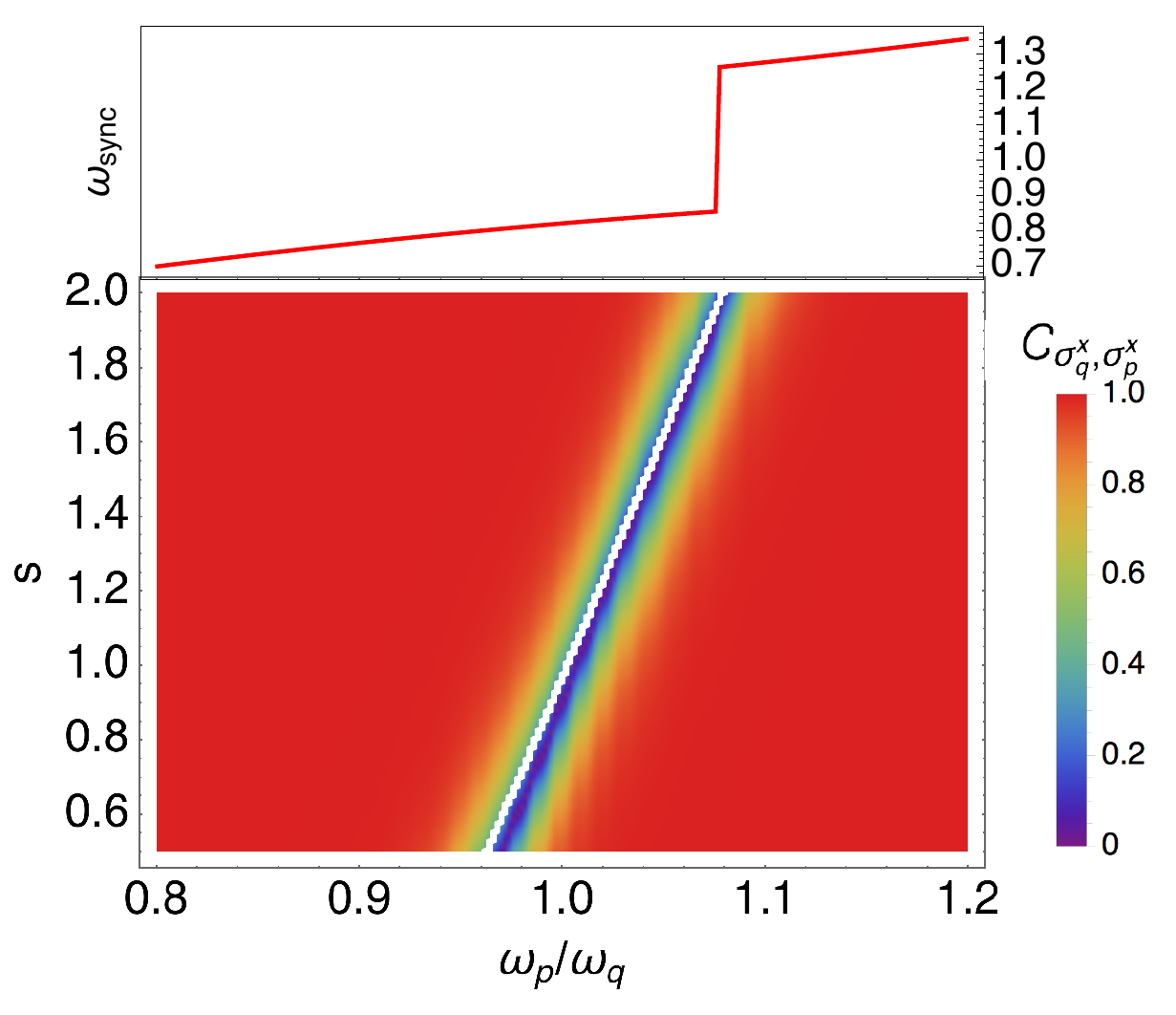}
\caption{(a) Synchronization frequency as a function of $\omega_p/\omega_q$ for $s=2$ and 
(b) absolute value of ${\cal C}_{\sigma_p^x,\sigma_q^x}(t,\tau)$  as a function of $\omega_p/\omega_q$ and $s$
for $\lambda=0.2\,\omega_q$
and spectral density $J(\omega)=2 \,\gamma_0\, \omega^s \,\omega_c^2/(\omega_c^2+\omega^{2s})$.
The white line in (b) is the analytical solution of Eq. (\ref{s}). }
\label{figsync}
\end{figure}

We are now going to characterize a key feature for the probing scheme, namely, the
dependence of the transition between SS in phase and  antiphase on the environment features.
Let us start from the case of a power-law spectrum  $J(\omega)\sim \omega^s$ with a high-energy cutoff
(later, we will discuss  generic densities $J$). The condition for the absence of synchronization
$\tilde\gamma_1/ \tilde\gamma_2=1$ is satisfied along a 
line in the $\omega_p-s$ diagram which corresponds to 
\begin{equation}
\log_{ \bar E_1/ \bar E_2}\tan^2( \bar\theta_+ +  \bar\theta_-) = s,\label{s}
\end{equation}
where the bar indicates that all the parameters must be calculated at a given probe frequency $\omega_p=\bar{\omega}_p$.
Synchronization  ${\cal C}$ as evaluated from the dynamical equations is shown in Fig. \ref{figsync}(b) after a transient time. 
We see that it fits the analytic prediction of Eq. (\ref{s}) (white line) up to
a slight displacement 
due to the spectral cutoff.  A second key aspect for a probing
scheme is that the presence of SS manifests itself in the local dynamics 
of the probe.
As we have anticipated before, the sharp transition between in-phase and
anti-phase SS leads also  to a change in synchronization frequency.
This is a key feature represented in Fig. \ref{figsync}  that will be
exploited in the probing protocol in order to reconstruct the spectral density.

\begin{figure}[t]
\includegraphics[width=9cm]{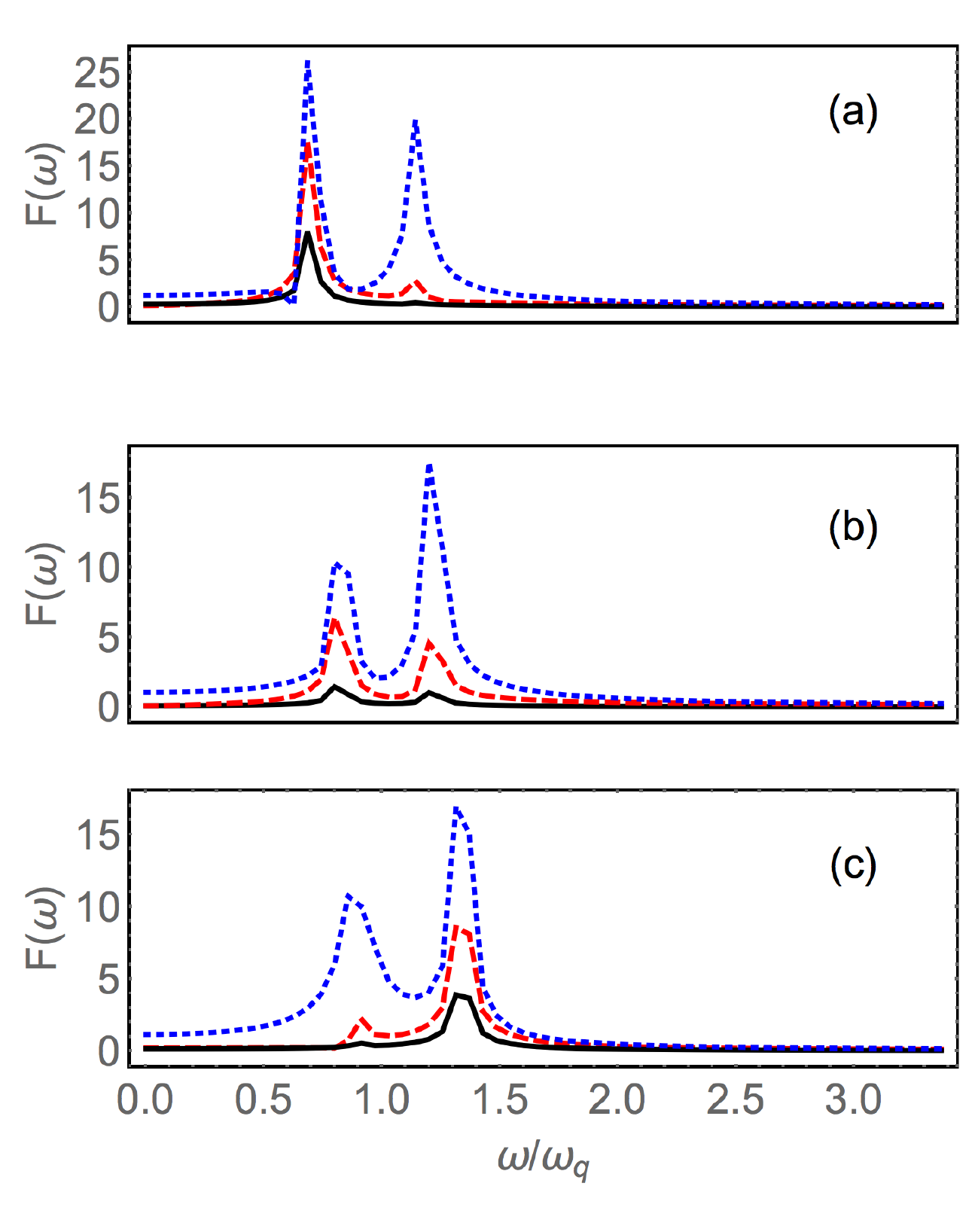}
\caption{Absolute value of the Fourier transform $F( \omega)$ of $\langle\sigma_p^x(t)\rangle$. The three panels refer to the three synchronization lines of Fig. \ref{fig2}: 
(a) $\omega_p/\omega_q=0.8$, (b) $\omega_p/\omega_q=1$, and (c) $\omega_p/\omega_q=1.2$. As in Fig. \ref{fig2}, the bath is Ohmic, and  $\lambda=0.2\, \omega_q $. The blue dotted
lines are taken calculating $F( \omega)$ during the time window $T_1=\{0,\,110 \,\omega_q\}$, the red dashed lines concern  the time window $T_2=\{ 100,\,210 \,\omega_q\}$,and, finally, the black 
solid lines refer to $T_3=\{200,\,310 \,\omega_q\}$.}
\label{figpeaks2}
\end{figure}

Assuming full control of the probe qubit ($\omega_p$ and the
coupling $\lambda $), we need first to infer the 
unknown system frequency $\omega_q$ from the probe dynamics. This can be
achieved either in the transient dynamics or after 
synchronization arises.
In the first case, the two eigenfrequencies of
$H_S$ can be measured locally from the probe dynamics, e.g., monitoring 
$\langle\sigma_p^x(t)\rangle$
in the initial transient time window. The frequency spectrum of the probe
is peaked around two values corresponding to $E_{1,2}$, from which
both $\omega_q$ and the spin-spin coupling $\lambda $ can be inferred.
Otherwise, after SS arises,   oscillations become
monochromatic and last during  a long transient: then the probe signal $\langle\sigma_p^x(t)\rangle$
can be measured on the two sides of the in-phase and antiphase synchronization.
With these two measurements at different probe frequencies, again, one can infer both 
$\omega_q$ and $\lambda$. Examples of probe frequency spectra are shown
in Fig. \ref{figpeaks2}, where we show the typical form of the absolute value of the Fourier transform $F( \omega)$ of $\langle\sigma_p^x(t)\rangle$ in the presence of in-phase SS, in the absence of SS and in the presence of antiphase SS. 
The spectra are obtained from the signal in different time windows and after an initial transient the three regimes are clearly recognized.
 During the first stage of the dissipative dynamics,  the presence of both the eigenmodes can be clearly observed. Afterwards,  in the case where 
synchronization is not expected to take over [Fig.  \ref{figpeaks2}(b)] the two peaks die with very similar time scales. In the other cases, after a transient, one of the peaks is still present, while the other has already disappeared.

\begin{figure}[t]
\includegraphics[width=8cm]{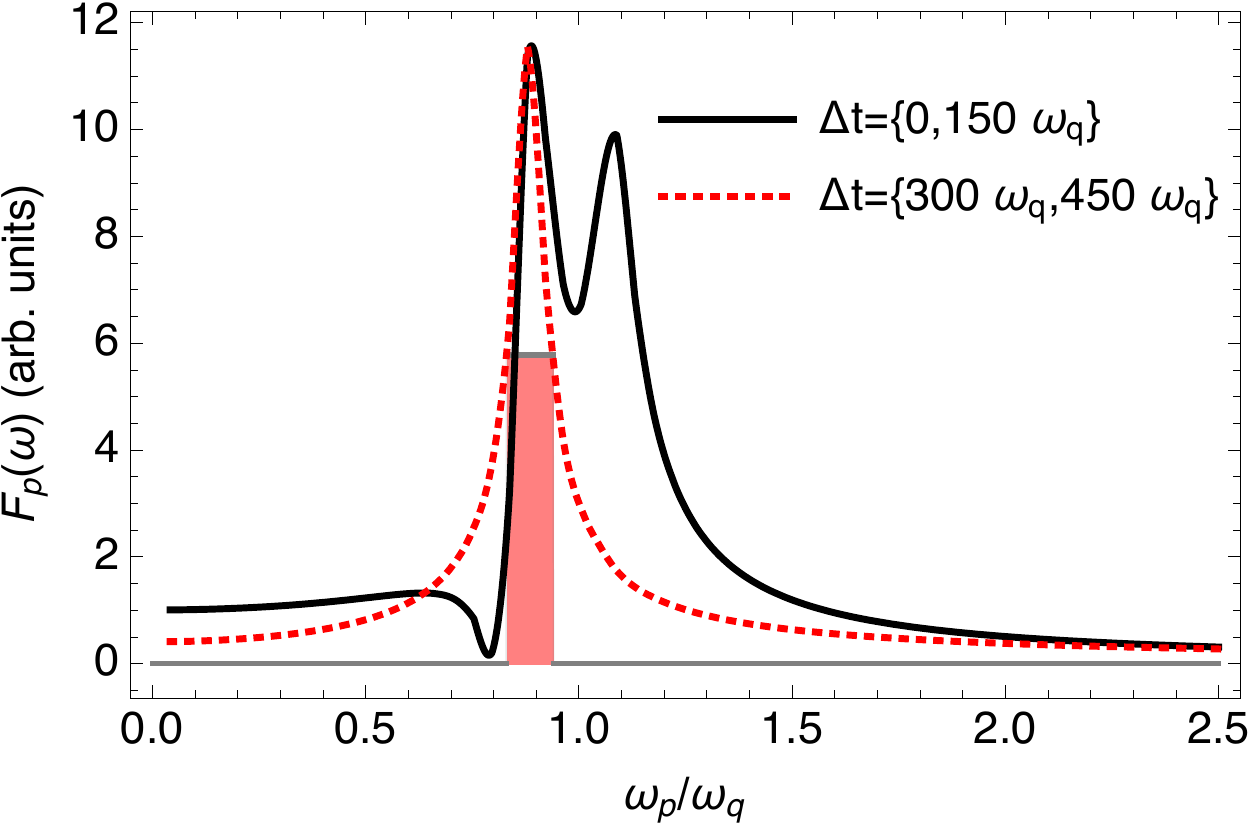}
\caption{Absolute value of the Fourier transform $F( \omega)$ of $\langle\sigma_p^x(t)\rangle$ in the initial transient (black solid line) and after SS has occurred (red dashed line).
In the last case the peak line-width can be determined. In the case simulated here the system-bath coupling is stronger than in the previous cases ($\gamma_0=2.5*10^{-2}\omega_q$).}
\label{figbroadpeaks}
\end{figure}

We have now all the elements for a protocol to reconstruct the whole profile $J(\omega)$ exploiting 
the critical behavior observed in  the passage from synchronization to
antisynchronization, across which a macroscopic frequency jump from $E_1$ to
$E_2$ takes place. We have seen that in the case of a
power-law spectrum (Fig. \ref{figsync}), finding the value  of the qubit energy split
$\bar\omega_p$ around which the discontinuity occurs allows us to determine the value of $s$ and then
infer the complete shape of $J(\omega)$, Eq. (\ref{s}). What about generic spectral densities?
In the absence of any prior
knowledge about the shape of the spectral density $J(\omega)$, the value $\bar\omega_p$
gives us the value of  the 
ratio $J(E_1)/J(E_2)$. In this case, a collection of different values of
$\bar\omega_p$ can be obtained by varying the system-probe coupling strength $\lambda$ and can then be used through numerical fit to complete the $full$ reconstruction
of $J(\omega)$.

An approach generally used to reconstruct the spectral density 
of a thermal bath consists of measuring the $linewidth$ of the frequency components of the dissipative system.
Here, however, we are discussing the case where  $q$ cannot be directly accessed.
Still, one can monitor the quantum \textit{probe} linewidth, relative, for instance, to the two-time correlation functions $\langle \sigma_+
(0)\sigma_- (\tau)\rangle_{ss}$ averaged over the steady state and then Fourier transform it \cite{gardiner}.
In this case the reconstruction of  $J(\omega)$ is limited by the precision in assessing 
such a FWHM in the frequency peaks. In fact, the lack of
a clear separation forbids a precise FWHM measurement of such peaks, making  the
precise estimation of the damping rates and the consequent reconstruction of $J(\omega)$ difficult.
 The advantage of our proposed scheme, instead, 
is that it is based on a switch detection: the measurement upon which the probing is based 
needs only to resolve a $sharp$ transition in the frequency of the oscillatory dynamics
when tuning the probe [Fig. \ref{figsync}(a)]. As a further consideration, SS can be of help also within 
standard probing strategies, as it would
filter out the fast-decaying  mode and make the line-width evaluation possible, as displayed in Fig. \ref{figbroadpeaks}.

\subsection{Comparison with other synchronization indicators}

With SS being the basis of this protocol, it is interesting to assess the role played by the indicator chosen to characterize it.
In Ref. \cite{ameri}, the asymptotic mutual information (MI) was proposed as an order
parameter to discriminate the presence of synchronization. However, this SS measure
is not of help in our case.
Indeed, its
failure is pretty evident for $T=0$, where the asymptotic state  
depends on only  the parameters of $H_S$, irrespective of the shape of $J(\omega)$.
Furthermore, the MI smoothly varies  only  as a function of 
$\omega_p$, without showing any signature either 
of the critical behavior reported in Fig. \ref{figsync}(a) or of the corresponding
absence of synchronization shown in Fig. \ref{figsync}(b). We have also checked the preasymptotic 
behavior of the MI without finding any relationship with synchronization.
SS is sometimes also associated with the presence of 
classical and quantum correlations. A similar approach was taken, for instance,
in Ref. \cite{zhu}, where synchronization was studied between clouds of dipoles
of different species ($a$ and $b$). There,  the quantity $\langle \sigma_+^a 
\sigma_-^b\rangle$, averaged over all  possible pairs of dipoles, was used to
infer the existence of a synchronized phase. Although  $\langle \sigma_+^{(q)}
\sigma_-^{(p)}\rangle$ does not need to coincide with $\cal{C}$,
we found  qualitative agreement between them (see Appendix \ref{appE}).

\section{Conclusions and outlook}

In this paper we have proposed a probing scheme for an out-of-equilibrium system (namely, a dissipative qubit) based on spontaneous synchronization. Possible implementations of such a scheme, consisting of coupled spins and bosons, even in the strong-coupling regime, are arrays of trapped ions \cite{ions} or setups with superconducting circuits and qubits \cite{circuitQED}. Quantum simulation of the  phase-antiphase transition could be observed by employing ultracold atoms in optical lattices \cite{lewenstein}, adding local dissipation to a two-site effective spin chain as proposed in Ref. \cite{schwager}.
 The problem of characterizing the dynamics of dissipative two-level fluctuators
of unclear physical origin  is especially important in
superconducting quantum bit circuits based on Josephson phase junctions \cite{simmonds}. 

An external qubit was used to probe the temperature  of such two-level systems in Ref. \cite{lisenfeld2010}. More recently, 
 their dissipative dynamics  was probed by monitoring the Josephson phase-qubit using standard interferometric techniques \cite{lisenfeld}.  
 The protocol presented in this paper would offer an alternative route to performing this kind of task  based on a simple switch measurement.

\acknowledgments

Funding from EU project QuProCS (Grant
Agreement No. 641277), MINECO and FEDER projects (Nomaq Grant
No. FIS2014-60343-P  and QuStruct Grant No. FIS2015-66860-P), and ``Vicerectorat d'Investigaci\'o  i Postgrau"  of the UIB is acknowledged.

\appendix

\section{Model}\label{appA}
Let us consider two interacting spins, whose free Hamiltonian reads
\begin{equation}
H_S=\frac{\omega_q}{2}\sigma_q^z+\frac{\omega_p}{2}\sigma_p^z+\lambda \sigma_q^x \sigma_p^x,\label{eq1}
\end{equation}
and assume that the qubit $q$ is immersed in a bosonic environment through 
\begin{equation}
H_I= \sum_k \gamma_k (a^{\dag}_k+ a_k)\sigma_q^x .
\end{equation}

$H_S$ can be diagonalized through the Jordan-Wigner transformation that, in the case of two qubits, reads
 \begin{eqnarray}
\sigma_q^z&=&1-2 c_1^\dag c_1,\\
\sigma_p^z&=&1-2 c_2^\dag c_2,\\
\sigma_q^x&=& c_1^\dag+ c_1,\\
\sigma_p^x&=& (1-2 c_1^\dag c_1) (c_2^\dag+ c_2).\label{s2x}
\end{eqnarray}
In the new fermionic space, the system Hamiltonian is 
\begin{equation}
H_S=\frac{\omega_q}{2}(1-2 c_1^\dag c_1)+\frac{\omega_p}{2}(1-2 c_2^\dag c_2)+\lambda \ (c_1^\dag- c_1) (c_2^\dag+ c_2).
\end{equation}
The diagonalization is then obtained by combining the Bogoliubov transformation
\begin{eqnarray}
c_1=\cos \theta_+ \xi_1+ \sin\theta_+ \xi_2^\dag ,\\ c_2=\cos \theta_+ \xi_2- \sin\theta_+ \xi_1^\dag ,
\end{eqnarray}
together with the rotation
\begin{eqnarray}
\xi_1=\cos\theta_-\eta_1^\dag+\sin\theta_-\eta_2^\dag,\\
\xi_2=\cos\theta_-\eta_2^\dag-\sin\theta_-\eta_1^\dag.
\end{eqnarray}
The conditions that bring $H_S$ to its diagonal form 
\begin{equation}
H_S=E_1(\eta_1^\dag \eta_1-1/2)+E_2 (\eta_2^\dag \eta_2-1/2)
\end{equation}
are 
\begin{equation}
\tan 2 \theta_+=\frac{2 \lambda}{\Delta}
\end{equation}
and 
\begin{equation}
\tan 2 \theta_-=\frac{2 \lambda}{\delta},
\end{equation}
where $\Delta=\sqrt{4 \lambda^2+(\omega_q+\omega_p)^2}$ and $\delta=\sqrt{4 \lambda^2+(\omega_q-\omega_p)^2}$. Finally, the single fermion energies are given by $E_1=(\Delta+\delta)/2$ and $E_2=(\Delta-\delta)/2$.

\section{Master equation}\label{appB}
In terms, of fermionic quasiparticles, the system operator $\sigma_q^x $ entering in $H_I$, written in the interaction picture with respect to $H_S$, admits the simple form 
\begin{eqnarray}
\tilde\sigma_q^x(t) &=&\cos (\theta_++\theta_-)(\eta_1^\dag e^{iE_1t}  + \eta_1 e^{-iE_1t})
\nonumber\\
&+&\sin(\theta_++\theta_-)(\eta_2^\dag e^{iE_2t}+ \eta_2e^{-iE_2t}).
\end{eqnarray}
Then, under the secular approximation, the dissipative part of the master equation will be simply given by the sum of terms in which each term corresponds to one of the frequencies $\pm E_i$  \cite{bp}: 
\begin{eqnarray}
{\cal D}(\rho)&=&\cos^2 (\theta_++\theta_-) \gamma(E_1)\left[\eta_1\rho\eta_1^\dag-\frac{1}{2}\{  \eta_1^\dag\eta_1,\rho\}\right]\nonumber\\
&+&\cos^2 (\theta_++\theta_-)\gamma(-E_1)\left[\eta_1^\dag\rho\eta_1-\frac{1}{2}\{  \eta_1\eta_1^\dag,\rho\}\right]\nonumber\\
&+&\sin^2 (\theta_++\theta_-)\gamma(E_2)\left[\eta_2\rho\eta_2^\dag-\frac{1}{2}\{  \eta_2^\dag\eta_2,\rho\}\right]\nonumber\\
&+&
\sin^2 (\theta_++\theta_-)\gamma(-E_2)\left[\eta_2^\dag\rho\eta_2-\frac{1}{2}\{  \eta_2\eta_2^\dag,\rho\}\right]\nonumber\\.\label{diss}
\end{eqnarray}  
Here, we have defined 
\begin{eqnarray}
\gamma(E_i) &=&2\pi J(E_i)[1+n(E_i)],\\
\gamma(-E_i) &=&2\pi J(E_i)n(E_i),
\end{eqnarray} 
 where $n(x)=1/(e^{x/T}-1)$ and $T$ is the temperature in units of the Boltzmann constant. 
 
It is useful to absorb  the trigonometric factors in Eq. (\ref{diss}) and define the decay rates  $\tilde\gamma_1^+=\cos^2 (\theta_++\theta_-)J( E_1)[1+n(E_1)]$, $\tilde\gamma_1^-=\cos^2 (\theta_++\theta_-)J( E_1)$, $\tilde\gamma_2^+=\sin^2 (\theta_++\theta_-)J( E_2)[1+n(E_2)]$, and $\tilde\gamma_2^-=\sin^2 (\theta_++\theta_-)J( E_2)$. Then,we can write a complete set of equations of motion for the density-matrix elements of the system in the fermionic basis. In  the interaction picture with respect to $H_S$  we have the following blocks of equations relevant for SS:
\begin{eqnarray}
\frac{d \rho_{00,01}}{dt}=\tilde\gamma_1^+  \rho_{10,11}  -\frac{1}{2}(2\tilde\gamma_1^-+\tilde\gamma_2^++\tilde\gamma_2^-)\rho_{00,01},\label{s1}\\
\frac{d \rho_{10,11}}{dt}=\tilde\gamma_1^-  \rho_{00,01}  -\frac{1}{2}(2\tilde\gamma_1^++\tilde\gamma_2^++\tilde\gamma_2^-)\rho_{10,11},\label{s2}
\\\frac{d \rho_{00,10}}{dt}=\tilde\gamma_2^+  \rho_{01,11}  -\frac{1}{2}(2\tilde\gamma_2^-+\tilde\gamma_1^++\tilde\gamma_1^-)\rho_{00,10},\label{s3}\\
\frac{d \rho_{01,11}}{dt}=\tilde\gamma_2^-  \rho_{00,10}  -\frac{1}{2}(2\tilde\gamma_2^++\tilde\gamma_1^++\tilde\gamma_1^-)\rho_{01,11},\label{s4}
\end{eqnarray}
together with their conjugate equations.

\section{Spontaneous synchronization}\label{appC}

We need to study the evolution of
\begin{eqnarray}
\langle\sigma_q^x(t)\rangle&=&\cos (\theta_++\theta_-){\rm Tr}[(\eta_1^\dag   + \eta_1)\rho(t)]\nonumber\\ &+&\sin(\theta_++\theta_-){\rm Tr}[(\eta_2^\dag + \eta_2)\rho(t)]
\end{eqnarray}
and 
\begin{eqnarray}
\langle\sigma_p^x(t)\rangle&=&-\cos (\theta_+-\theta_-){\rm Tr}[ {\cal P}(\eta_2^\dag- \eta_2)\rho(t)]\nonumber\\
&-&\sin (\theta_+-\theta_-){\rm Tr}[ {\cal P}(\eta_1^\dag- \eta_1)\rho(t)],
\end{eqnarray}
where  ${\cal P}=(1-2 \eta_1^\dag \eta_1)(1-2 \eta_2^\dag \eta_2)$ is the parity operator.

\begin{figure}
\includegraphics[width=8cm]{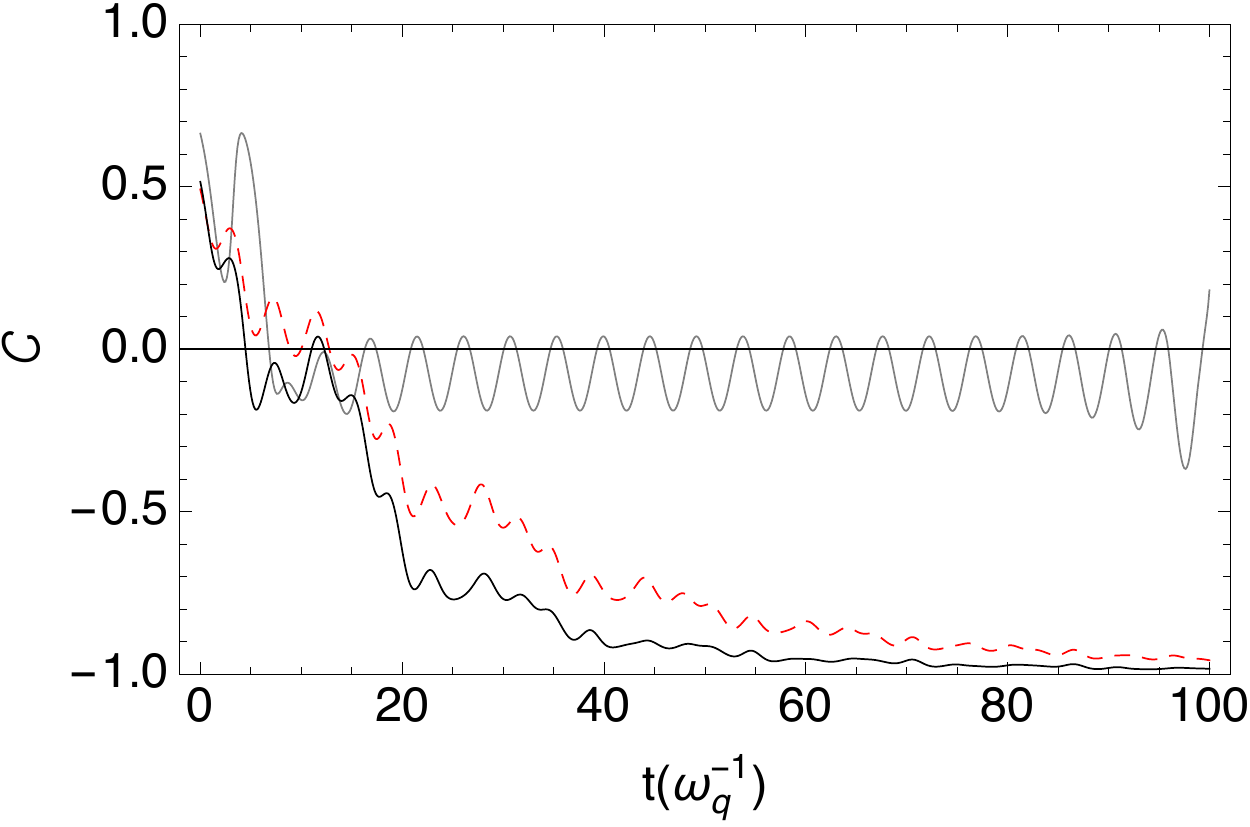}
\caption{Dynamical synchronization as a function of the temperature. The probe frequency is set to  $\omega_p/\omega_q=0.8$ (see Fig. \ref{fig1}) for the initial state and the bath. The red (dashed) line is the one obtained at $T=0$. It is compared with the case where  $T=1$ (black line) and with $T=10$ (gray solid line). As we can notice, in principle, the finite-temperature regime does not necessarily prevents the system from becoming synchronized. However, if $T$ is too high, the system reaches the effective stationary state before synchronization can take place. To give a rough estimation, considering a spin-spin coupling around $\simeq 100 {\rm MHz}$ (see, for instance, Ref. \cite{viehmann}), $T=1$ (in our units) would correspond to a few dozen  $\mu K$.  }\label{figtemp}
\end{figure}

We have

\begin{eqnarray*}
\langle\eta_1^\dag+ \eta_1\rangle &=&(\rho_{00,10}+\rho_{10,00})+(\rho_{01,11}+\rho_{11,01}),\\
\langle\eta_2^\dag+ \eta_2 \rangle &=&(\rho_{00,01}+\rho_{01,00})+(\rho_{10,11}+\rho_{11,10}),\\
\langle {\cal P}(\eta_1^\dag- \eta_1)\rangle &=&-(\rho_{00,10}+\rho_{10,00})+(\rho_{01,11}+\rho_{11,01}),\\
\langle {\cal P}(\eta_2^\dag- \eta_2)\rangle &=&-(\rho_{00,01}+\rho_{01,00})+(\rho_{10,11}+\rho_{11,10}).\end{eqnarray*}
Considering, for instance, Eqs. (\ref{s1}) and (\ref{s2}), the two decay rates are $(\tilde\gamma_2^++\tilde\gamma_2^-)/2$ and $(\tilde\gamma_2^++\tilde\gamma_2^-)/2+(\tilde\gamma_1^++\tilde\gamma_1^-)$. Then, it is immediately possible to find the slowest one. 
After the first transient, the evolution of these matrix elements (now we also consider, apart from the negligible Lamb shift, the Hamiltonian part of the evolution) is given by 
 \begin{eqnarray}
 \rho_{00,01}(t)&\sim &  e^{i E_2 t }e^{-(\tilde\gamma_2^-+\tilde\gamma_2^+)t/2}\nonumber\\&\times& \frac{\gamma_1^+}{\gamma_1^-+\gamma_1^+}[\rho_{10,11}(0)+\rho_{00,01}(0) ],\\
 \rho_{10,11}(t)&\sim & e^{i E_2 t }e^{-(\tilde\gamma_2^-+\tilde\gamma_2^+)t/2}\nonumber\\&\times&\frac{\gamma_1^-}{\gamma_1^-+\gamma_1^+}[\rho_{10,11}(0)+\rho_{00,01}(0) ],
 \end{eqnarray}
 and 
 \begin{eqnarray}
 \rho_{00,10}(t)&\sim & e^{i E_1 t }e^{-(\tilde\gamma_1^-+\tilde\gamma_1^+)t/2}\nonumber\\&\times&\frac{\gamma_2^+}{\gamma_2^-+\gamma_2^+}[\rho_{01,11}(0)+\rho_{00,10}(0) ],\\
 \rho_{01,11}(t)&\sim & e^{i E_1 t }e^{-(\tilde\gamma_1^-+\tilde\gamma_1^+)t/2}\nonumber\\&\times&\frac{\gamma_2^-}{\gamma_2^-+\gamma_2^+}[\rho_{01,11}(0)+\rho_{00,10}(0) ].
 \end{eqnarray}
 Combining these elements, we have
\begin{eqnarray}
\langle\sigma_q^x(t)\rangle &\sim & 2 \cos( \theta_+-\theta_-)e^{-(\tilde\gamma_1^-+\tilde\gamma_1^+)t/2}\nonumber\\
&\times& {\rm Re}\{e^{i E_1 t }[\rho_{01,11}(0)+\rho_{11,10}(0)] \}
\nonumber\\
&+&2\sin(\theta_--\theta_-)e^{-(\tilde\gamma_2^-+\tilde\gamma_2^+)t/2}
\nonumber\\&\times&{\rm Re}\{ e^{i E_2 t }[\rho_{10,11}(0)+\rho_{00,01}(0)] \}
\end{eqnarray}
 and
\begin{eqnarray}
\langle\sigma_p^x(t)\rangle&\sim & 2 \frac{\sin( \theta_+-\theta_-)}{\coth[E_2/(2 T)]}e^{-(\tilde\gamma_1^-+\tilde\gamma_1^+)t/2} \nonumber\\
&\times&{\rm Re}\{e^{i E_1 t }[\rho_{01,11}(0)+\rho_{11,10}(0)]  \}
\nonumber\\
&+&2\frac{\cos( \theta_+-\theta_-)}{\coth[E_1/(2 T)]}e^{-(\tilde\gamma_2^-+\tilde\gamma_2^+)t/2}
\nonumber\\
&\times&{\rm Re}\{ e^{i E_2 t }[\rho_{10,11}(0)+\rho_{00,01}(0)] \}.
\end{eqnarray}
The zero-temperature limit of these expressions is given in the main text.
The temperature effect over synchronization is illustrated in Fig. \ref{figtemp}

\section{Synchronization diagram}
\begin{figure}
\includegraphics[width=9cm]{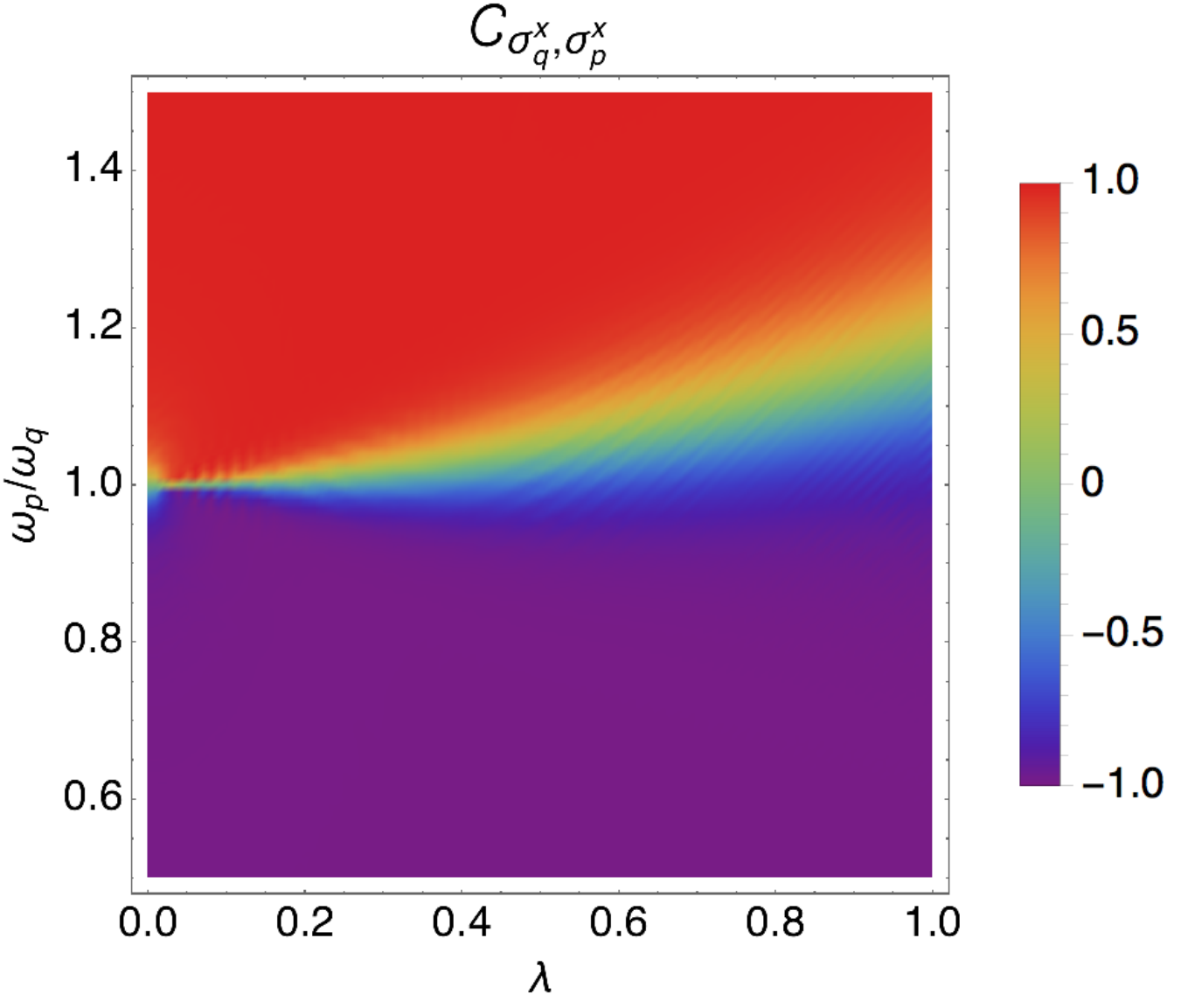}
\caption{Synchronization diagram as a function of the ratio $\omega_p/\omega_q$ and of the coupling $\lambda$. The synchronization has been measured at $t=350\omega_q^{-1}$ assuming an Ohmic bath with a frequency cutoff  $\omega_c=20\omega_q $. }\label{fig2}
\end{figure}

Unlike the case of a common bath \cite{PRAsync,syncNET,Giorgi}, here, the emergence of synchronization is favored by a strong detuning between the dissipative qubit and the probe. This is illustrated for the case of an Ohmic environment in Fig. \ref{fig2}.
This result can be explained by observing that the system operator interacting with the bath $\sigma_q^{x}$ can  induce two kinds of transitions (at frequencies $E_i$) between the eigenmodes. In the presence of strong detuning, the matrix elements corresponding to the two transitions have highly unbalanced weights. 
This makes the whole dynamics slower, but the reason the system achieves synchronization relies on the fact that the decay rates are related to the square of such coupling coefficients [see Eq. (\ref{diss})]. Figure \ref{fig2} also shows that the synchronization-antisynchronization transition gets less sharper for moderately strong values of $\lambda$. 

 \section{Synchronization and spin-spin correlations}\label{appE}
\begin{figure}[H]
\includegraphics[width=8cm]{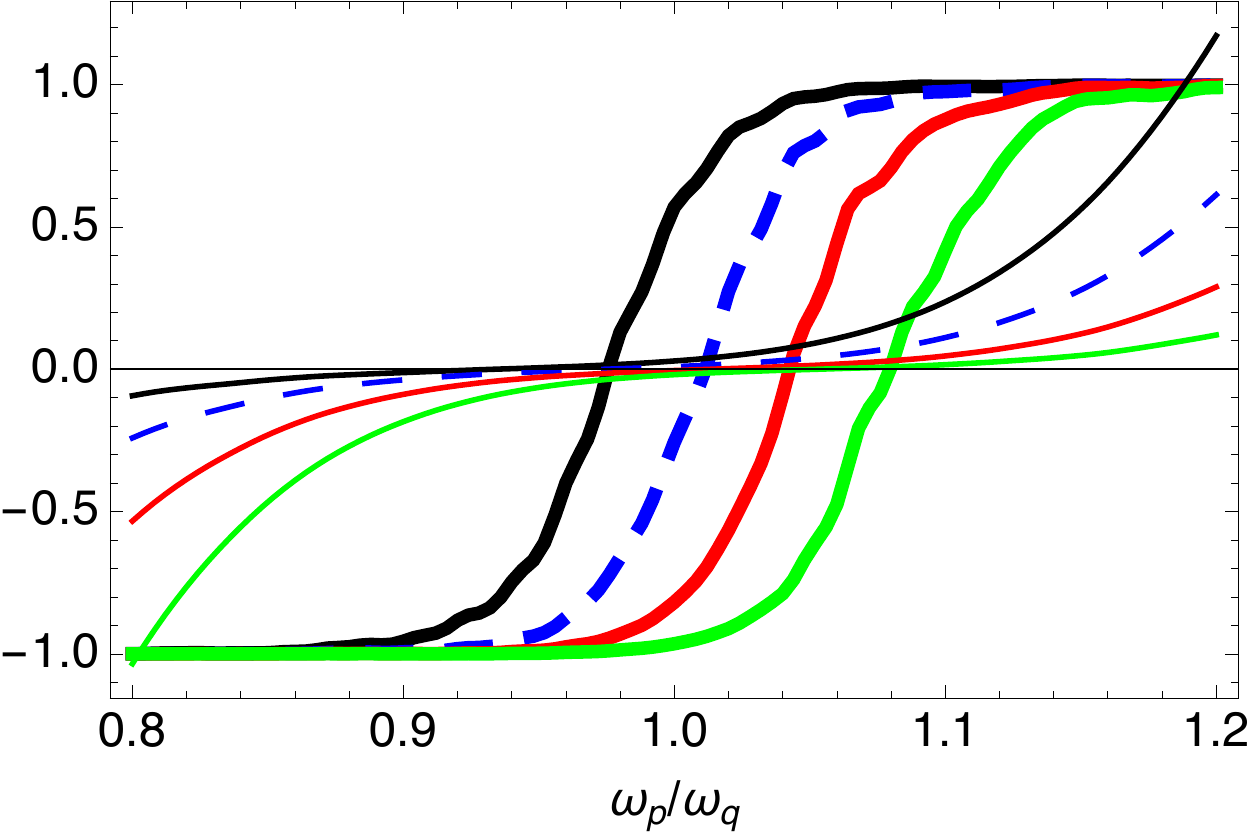}
\caption{Comparison between $\cal{C}$ (thick lines) and $10^2 \langle \sigma_+^{(q)}(t)\sigma_-^{(p)}(t)\rangle$ (thin lines). The environment is assumed to have the power-law spectral density $J(\omega)\sim \omega^s$ away from the frequency cutoff. The colors correspond to different values of $s$: $s=0.5$ (black lines),   $s=1$ [blue (gray) dashed lines],   $s=1.5$ [red
(dark gray) solid lines], and  $s=2$ [green (light gray) solid lines]. The qubit-probe coupling is set to $\lambda=0.2 \omega_q$, and all the data are taken at $t=300 \omega_q$.}
\label{figcorr}
\end{figure}

 As suggested in Ref. \cite{zhu}, a reliable measure of synchronization of two spins (or families of spins) $q$ and $p$ is given by the correlation $\langle \sigma_+^{(q)}\sigma_-^{(p)}\rangle$, which quantifies the phase locking between them. In our model, such an indicator is related to a set of equations of motion autonomous with respect to    Eqs. (\ref{s1})-(\ref{s4}):
  \begin{eqnarray}
 \frac{d \rho_{00,00}}{dt}=\tilde\gamma_1^+\rho_{10,10}+\tilde\gamma_2^+\rho_{01,01}-(\tilde\gamma_1^-+\tilde\gamma_2^-)\rho_{00,00},\nonumber\\
  \frac{d \rho_{01,01}}{dt}=\tilde\gamma_1^+\rho_{11,11}+\tilde\gamma_2^-\rho_{00,00}-(\tilde\gamma_1^-+\tilde\gamma_2^+)\rho_{01,01},\nonumber\\
   \frac{d \rho_{10,10}}{dt}=\tilde\gamma_2^+\rho_{11,11}+\tilde\gamma_1^-\rho_{00,00}-(\tilde\gamma_1^++\tilde\gamma_2^-)\rho_{10,10},\nonumber\\
   \frac{d \rho_{11,11}}{dt}=\tilde\gamma_1^-\rho_{01,01}+\tilde\gamma_2^-\rho_{10,10}-(\tilde\gamma_1^++\tilde\gamma_2^+)\rho_{11,11},
 \end{eqnarray}
together with
\begin{eqnarray}
\frac{d \rho_{00,11}}{dt}=-\frac{1}{2}(\tilde\gamma_1^++\tilde\gamma_1^-+\tilde\gamma_2^++\tilde\gamma_2^-)\rho_{00,11},\nonumber\\
\frac{d \rho_{11,00}}{dt}=-\frac{1}{2}(\tilde\gamma_1^++\tilde\gamma_1^-+\tilde\gamma_2^++\tilde\gamma_2^-)\rho_{11,00},\nonumber\\
\frac{d \rho_{01,10}}{dt}=-\frac{1}{2}(\tilde\gamma_1^++\tilde\gamma_1^-+\tilde\gamma_2^++\tilde\gamma_2^-)\rho_{01,10},\nonumber\\
\frac{d \rho_{10,01}}{dt}=-\frac{1}{2}(\tilde\gamma_1^++\tilde\gamma_1^-+\tilde\gamma_2^++\tilde\gamma_2^-)\rho_{10,01}.
\end{eqnarray}
 Actually, a tight relationship between the sets of solutions of the two problems can be expected as together they describe the dynamics of a density matrix. Indeed, in Fig. \ref{figcorr} we compare  $\cal{C}$ to $\langle \sigma_+^{(q)}(t)\sigma_-^{(p)}(t)\rangle$ in the long-time regime and observe that when the system is synchronized, the spin-spin correlations are remarkably stronger than when there is no synchronization. This happens irrespective of the type of environment considered.
 \newline
 
\begin{thebibliography}{10}
\bibitem{footnote} It is worth noting that a synchronous dynamics
can also be observed  without dissipation, if $\omega_q=\omega_p$
and if the initial state is symmetrical. This trivial case will not be discussed here.
\end{thebibliography}
\end{document}